\documentclass{elsart}
\usepackage{graphicx}
\usepackage{amssymb}
\begin{document}

\begin{frontmatter}

 \title{{Exchange parameters in Fe-based molecular magnets}}
 \author{A. V. Postnikov\corauthref{cor1}}
 \corauth[cor1]{Corresponding author. Tel.: +49-541-9692377; fax: -9692351.}
 \ead{apostnik@uos.de}
 \address{Universit\"at Osnabr\"uck -- Fachbereich Physik,
          D-49069 Osnabr\"uck \thanksref{label3}}
 \thanks[label3]{Permanent address: %
                 Institute of Metal Physics, 620219 Yekaterinburg, Russia}
 \author{G. Bihlmayer and S. Bl\"ugel}
 \address{Institut f\"ur Festk\"orperforschung, Forschungszentrum J\"ulich,
 D-52425 J\"ulich, Germany}

\begin{abstract}
The calculation of interatomic magnetic exchange interactions entering 
the Heisenberg model from the standpoint of the density functional theory 
(DFT) is outlined for
two Fe-based molecular magnets: a trinuclear complex with a Schiff base ligand,
which makes an antiferromagnetically coupled frustrated system, and 
a model bipyrimidine-connected planar network of Fe ions. 
First-principles electronic structure calculations are performed
using the real-space method {\sc Siesta} and the full-potential linearized
augmented plane wave FLAPW method {\sc FLEUR},
correspondingly. We discuss the application of fixed spin moment technique 
for preparing the system in a given magnetic configuration, and 
the effect of intraatomic Coulomb correlation, approximated by the
LDA+$U$ technique, on the values of interaction parameters.
\end{abstract}


\end{frontmatter}

\section{Introduction}
\label{sec:intro}
Since the first systematizations of their chemistry
and basic properties by, e.g., Kahn in 1993 \cite{Kahn-book},
molecular magnets
are now developing into a promising class of magnetic
nanomaterials with a great  potential in applications in areas such as 
magnetic storage, quantum computing, or magneto-optical devices. 
An up-to-date review on microscopic properties (spin density, exchange 
interactions, magnetic anisotropy) of molecular magnets
accessible in experiment and from first-principles calculations, 
is going to appear in Ref.~\cite{MolMag-book}. Its major part is summarized
as an internet publication \cite{Psik-highlight}.
In the present work we address the issue of extracting 
the interatomic exchange interaction parameters  entering
the Heisenberg model form the  density functional theory (DFT).
We discuss two chemical systems of different complexity. 
The actual calculations are done by two different methods, 
{\sc Siesta} \cite{siesta} using compact and strictly confined 
atom-centered basis functions (see Ref.~\cite{JPCM14-2745} for details), 
and the {\sc FLEUR} \cite{fleur} code, a realization of the highly accurate 
full-potential
linearized augmented plane wave (FLAPW) method. Both calculations used 
the generalized gradient approximation to the exchange-correlation functionals.

We skip the subtleties in the formulation of exchange parameters within 
the DFT; a detailed discussion can be found in Ref.~\cite{Psik-highlight}.
Basically, we refer to total energies in different specially prepared
magnetic configurations, calculated from first principles. 
Sec.~\ref{sec:Fe3} covers the extraction of meaningful results 
for a Fe-trinuclear system by applying the fixed spin moment formalism. 
Sec.~\ref{sec:Fe2} addresses a model Fe-binuclear system and, specifically, 
the importance of intraatomic correlation, treated beyond the conventional 
DFT formalism. 

\section{Fixed Spin Moment treatment of a Fe-trinuclear system}
\label{sec:Fe3}
One of the ways to arrive at the values of interaction parameters
in a magnetic system whose properties are well decribed by the Heisenberg 
model is
to compare total energies for different magnetic configurations. This is
rather straightforward if, say, ferromagnetic (FM) and antiferromagnetic
(AFM) states are well defined and at least metastable. A more complicated case
is a frustrated system, an example of which is made of three Fe atoms
incorporated into organic molecule. We consider specifically 
Fe$_3$(OAc)$_3$L$_3$, a synthesized by Boskovic 
\emph{et al.} \cite{InCh43-5053} derivate of the Schiff base 
H$_2$L = salicylidene-2-ethanolamine.
The molecular unit used in the electronic structure calculation 
by the {\sc Siesta} method is shown in Fig.~2 (a) of 
Ref.~\cite{InCh43-5053}. The magnetization measurements
reported in Ref.~\cite{InCh43-5053} indicate an (almost AFM) ground
state with the total spin $S$=1/2, coming about from the interaction 
of three $s$=5/2 spins of Fe(III) ions. Experimentally, the interaction 
parameters $J$ of 
the Heisenberg model, introduced in Ref.~\cite{InCh43-5053} as 
$H= -2\!\!\sum\limits_{\mbox{\tiny Fe pairs}} J_{ij}\,{\bf s}_i\,{\bf s}_j$,
were reported to be necessarily different for an acceptable fitting
of data, even as the molecule nearly maintains a trifold symmetry axis: 
$J_{ij}$ = $-$15.1 K, $-$13.6 K, and $-$12.4 K.

We proposed earlier \cite{EMRS-Fewheel} to use the Fixed Spin Moment (FSM)
method \cite{JPF14-L129} for selecting a magnetic configuration
of a molecular magnet (a six-center ``ferric wheel''), improving 
simultaneously the stability of the electronic structure calculation. 
For the present system this idea becomes even more important, because it is 
no more obvious how to simulate a (nearly) AFM trial magnetic configuration
in the DFT run, nor how to enforce an inversion of a selected
local magnetic moment. On the contrary, by running a sequence of FSM
calculations from the maximal nominal value of the total spin $S$=15/2
(and higher) down to zero, one can scan the variety of all possible
magnetic solutions within the DFT.

\begin{figure}
\centerline{\includegraphics*[width=\linewidth]{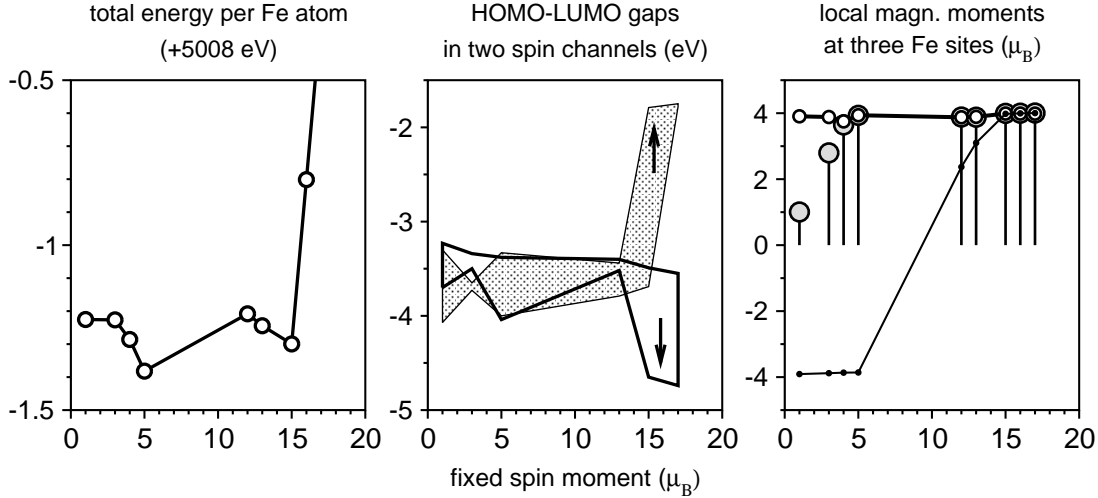}}
\caption{Results of the FSM calculation for [Fe$_3$(OAc)$_3$L$_3$].
See main text and explanations \cite{comment_FSM} for details.}
\label{fig:Fe3_FSM}
\end{figure}

The results of our FSM calculation are shown in Fig.~\ref{fig:Fe3_FSM}.
The total energy (left panel) has several local minima, the lowest one 
being for the FSM value of 5 $\mu_{\mbox{\tiny B}}$.
The right panel shows that this situation corresponds
to two Fe magnetic moments aligned in parallel and the third one
aligned in the opposite direction. A competitive energy minimum is found for 
FSM = 15 $\mu_{\mbox{\tiny B}}$, i.e., the ferromagnetic configuration.
The solution with FSM = 1 $\mu_{\mbox{\tiny B}}$ has higher energy than 
the latter one by 74 meV, per Fe atom. It should be noted that, whereas 
the ground state of the system is experimentally found to have $S$=1/2, 
it must have a mixed nature over many quantum states, and hence 
does not correspond to any single-determinant DFT result.
Nevertheless, the tendency for antiparallel coupling of Fe magnetic moments
is correctly provided by our calculation; moreover the order of magnitude
of the interaction constants can be correctly recovered: consistently
with the above formula for the Heisenberg Hamiltonian,
$-J \approx (E^{\uparrow\uparrow\uparrow}-E^{\uparrow\uparrow\downarrow})/
(8s^2)$ amounts to 57 K. The error (overestimation by a factor of about 4) 
in the DFT calculation is nearly the same as we found earlier for the
``ferric wheel'' \cite{EMRS-Fewheel}. A possible reason for
such systematic discrepancy can be the  underestimation of the intraatomic
Coulomb correlation, to be discussed in the next Section. 

Two issues might need further discussion. The HOMO-LUMO gaps shown
in the middle panel of Fig.~\ref{fig:Fe3_FSM} are different in the FSM
formalism for two spin directions, because the fixing of the total spin
amounts to an imposition of an external magnetic field, moving apart
the chemical potentials in two spin channels. Normally this would require
the consideration of a  Zeeman term when dealing with total energies.
However, for a system with a \emph{common} band gap in both spin channels
a unique chemical potential can be found for both spin directions,
therefore the FSM scheme would simply fix one or another of the
metastable magnetic configurations at no additional energy cost.
One can see that, indeed, a common band gap can be found in all situations
relevant for our discussion, i.e.\ with FSM = 1, 5, and 15 $\mu_{\mbox{\tiny B}}$,
thus justifying the extraction of the $J$ value above. Another observation 
concerns the local magnetic moments at Fe sites: as the FSM value
is changing, and even as the saturation of magnetization 
(at FSM = 15 $\mu_{\mbox{\tiny B}}$) is achieved, the local Fe magnetic
moments merely flip, essentially maintaining their magnitudes 
of 4 $\mu_{\mbox{\tiny B}}$, and \emph{not} 5 $\mu_{\mbox{\tiny B}}$.
The nominal value of $s$=5/2, associated
with a single Fe atom, comes about due to magnetic polarization of ligands.
This situation is identical to that described earlier for
``ferric wheels'' \cite{Psik-highlight,EMRS-Fewheel,Bedlewo-Fewheel}.
In spite of the very different structure of organic ligands, both systems
have similarities in the nearest neighbourhood and in the charge state
of Fe atoms (O$_6$ twisted octahedral coordination in ``ferric wheels''
vs. slightly distorted octahedral O$_5$N coordination in the present system).

\section{Fe-binuclear system: Effect of intraatomic correlation}
\label{sec:Fe2}
\begin{figure}[ht]
\centerline{\includegraphics*[width=0.9\linewidth]{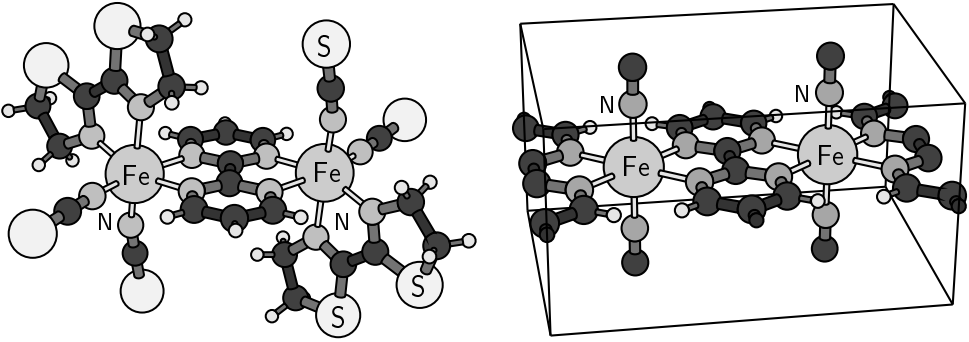}}
\caption{Molecular unit of the Fe-binuclear system
Fe[(bt)(NCS)$_2$]$_2$bpym described in Ref.~\cite{JAmChSoc114-4650}
(left) and a simplified model thereof, used in the calculation
(right). The Fe atom is in the middle of a slightly distorted N$_6$ octahedron.}
\label{fig:Fe2_molec}
\end{figure}

We turn now to the discussion of the effect of  on-site \emph{intraatomic} 
Coulomb correlation on \emph{interatomic} exchange parameters.
Our objective was to find a compact and yet realistic system, 
for which an electronic structure calculation could be done with the use
of an ultimately accurate method within the DFT, that is, the FLAPW
method, and to apply a phenomenological correction
by the LDA+$U$ formalism \cite{LDA+U} on top of it.
Specifically, we apply the realization of the method in the {\sc FLEUR}
code \cite{fleur}.
We derive a simple model system from the Fe-binuclear
complex described by Real \emph{et al.} \cite{JAmChSoc114-4650} being subject
to previous studies \cite{JAmChSoc121-10630}.
We neglect for the moment the spin crossover property, which makes
this system particularly interesting, and make our calculation
for the high-spin configuration on both Fe centers. The chemical formula of
the substance is [Fe(bt)(NCS)$_2$]$_2$-bpym, with bt = 2,$2'$-bithazoline
and bpym = 2,$2'$-bipyrimidine, and the molecular unit as shown
on the left of Fig.~\ref{fig:Fe2_molec}. We also performed electronic
structure calculation for this substance in its molecular and 
crystallized form by the {\sc Siesta} method, allowing structure
relaxation, the results of this study will be reported elsewhere.
For the present simulation, as the molecule is yet too large and too
low-symmetric for an affordable FLAPW calculation, we simplify the molecular
units, substituting the bt fragments by bpym -- that is realistic, as this is
encountered in related compounds -- see, e.g., \cite{JAmChSoc121-10630}.
Moreover we cut and connect the (NCS) tails into --N=C=N-- chains
that makes a dense 3-dimensional lattice out of planar 
Fe--bipyrimidine groups (Fig.~\ref{fig:Fe2_molec}, right panel).
The nearest neighbourhood of each Fe center remains a slightly distorted
N$_6$ octahedron, as was the case in the original molecular unit.
An idealization is that all N--Fe--N bonds make now right angles, 
and the repeated bipyrimidine fragments are coplanar.

\begin{figure}[hb]
\centerline{\includegraphics*[width=0.95\linewidth]{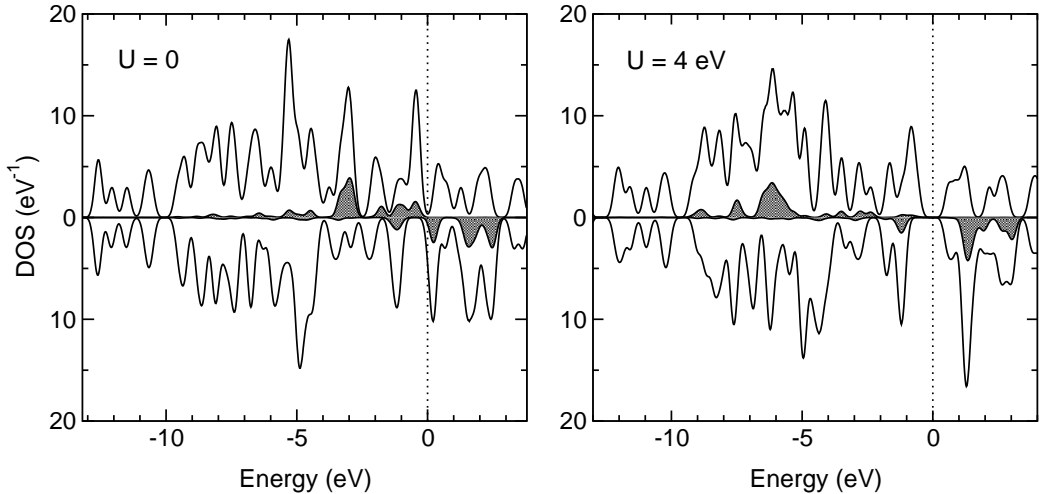}}
\caption{Total density of states per unit cell (outer solid line) and local DOS
per Fe site
(evaluated within the muffin-tin sphere of 2.3 Bohr; shown by
filled black) densities of states calculated for the model 
Fe-binuclear system, shown in the right panel of Fig.~\ref{fig:Fe2_molec},
for parallel orientation of Fe magnetic moments. 
Left panel: LDA calculation, right panel: LDA+$U$ calculation with $U$=4 eV.}
\label{fig:Fe2_dos}
\end{figure}

The calculated densities of states (per unit cell and per one Fe site)
in Fig.~\ref{fig:Fe2_dos} clearly indicate a hybridization of Fe$3d$ with 
the $2p$ states of neighbouring nitrogen atoms. Since the minority-spin 
Fe $3d$ states are not completely empty, the local magnetic moment is less 
than 5 $\mu_{\mbox{\tiny B}}$. As for the previous trinuclear system, 
the magnetic moment induced on nitrogen neighbours
substantially contributes to the net value, associated to each Fe site.
This resolves, at least in part, a controversy between the spin value $s$=5/2, 
expected at the Fe site from formal valence considerations, and the local 
magnetic moment of our calculation. The cumulative magnetic moment,
distributed over the Fe site and its neighbours, behaves like a ``rigid'' spin 
in the sense of the Heisenberg model. However, one can expect that a stronger
localization of this distributed magnetic moment at the Fe site
will be enforced by an inclusion of intraatomic Coulomb correlation,
underestimated in conventional DFT calculations. In order to check 
this possible effect, we brought in an additional Coulomb correlation
applying a semi-empirical ``LDA+$U$'' scheme \cite{LDA+U}. We choose 
the Coulomb interaction constant $U$ = 4 eV, which seems reasonable
for correlated Fe compounds \cite{PRB55-12822,JPCM11-2341}. One sees in the DOS 
of Fig.~\ref{fig:Fe2_dos} (right panel) the known effect of the ``+$U$''
correction to lower the occupied states in energy whereas shifting
unoccupied ones upwards. 

The values of magnetic moments and exchange parameters $J$, 
determined by the formula $E = -J\,{\bf s}_1 {\bf s}_2$, 
are listed in the Table. It is noteworthy that, whereas 
the total magnetic moment per Fe site is increased in the LDA+$U$
calculation, the \emph{local} Fe magnetic moment still does not exceed
4 $\mu_{\mbox{\tiny B}}$. On can conclude that
the spatial distribution of magnetic density is not confined
to the Fe atom even as the intraatomic Coulomb correlation is enhanced;
instead, the magnetization maintains its complex spatial distribution, 
spilling onto neighbouring N atoms. 

\begin{table}
\begin{center}
\caption{Local magnetic moments at the Fe site (in $\mu_{\mbox{\tiny B}}$, 
within muffin-tin sphere of 2.3 Bohr); total magnetic moments \emph{per} 
Fe center, total energy difference between FM and AFM configurations, 
and corresponding extracted values of the Heisenberg exchange parameter $J$,
assuming nominal spin value $s$=5/2.
}
\medskip
\begin{tabular}{|cccccc|}
 \hline
 & \multicolumn{2}{c}{$M_{\mbox{Fe}}$} & $M$/Fe & & \\ 
 \cline{2-3} 
 $U$ (eV) & FM & AFM & FM & 
 $E_{\mbox{\tiny FM}}-E_{\mbox{\tiny AFM}}$ (meV) & $J$ (K) \\
 \hline
  0   & 3.62 & 3.61 & 4.10 & 102.5 & $-$95  \\
  4   & 3.93 & 3.92 & 4.94 & ~76.8 & $-$71  \\
 \hline
\end{tabular}
\end{center}
\end{table}

As the intraatomic Coulomb correlation shifts apart on the energy scale
the centers of gravity of occupied and unoccupied states, it reduces
the interatomic exchange parameter, because the latter (in one of its
formulations, see \cite{PRB52-R5467}) has corresponding energy differences 
in the denominator. The same trend prevails in our case even as we estimate 
the values of $J$ by comparing the total energies of two magnetic 
configurations. The effect of Coulomb correlation, included on top of 
a straightforward DFT calculation, is therefore always towards lowering
the absolute values of Heisenberg exchange parameters. The same conclusion
has been drawn by Boukhvalov \emph{et al.} in their calculations for
``Mn$_{12}$'' \cite{PRB65-184435} and ``V$_{15}$'' \cite{Boulhvalov-V15}
molecular magnets. Comparing with the results
of the previous Section, one can conclude that the estimation of
exchange parameters in molecular magnets, that is already qualitatively correct 
(in sign and order of magnitude) in a conventional DFT calculation,
can be brought closer to the experimental estimates by reducing their 
numerical values due to additionally included intraatomic Coulomb
correlation.

\section*{Acknowledgements}
The authors thank Deutsche Forschungsgemeinschaft for financial support
(Priority Program `Molecular Magnetism'). A.V.P. thanks Colette Boskovic 
and Jos\'e Real for providing crystallographic information on substances
synthesized by them, and Danil Boukhvalov for making accessible his
manuscript on V$_{15}$ prior to publication.


\begin{thebibliography}{10}
\bibitem{comment_FSM}
It should be explained that the Fe-trinuclear complex reported in
  Ref.~\cite{InCh43-5053} crystallizes with two molecular units per primitive
  cell, which are slightly different. In particular, there is a formal
  disproportionality in the nominal electron number between these 
  two fragments. As
  a consequence, a single molecular unit we took for our calculation possesses
  an odd number of electrons. Therefore only odd FSM values correspond to
  integer number of electrons in both majority- and minority-spin channels and
  may lead to an appearance of HOMO-LUMO gaps, shown in the middle panel of
  Fig.~\ref{fig:Fe3_FSM}. However, the even FSM values can be formally applied 
  in the calculation as well, notwithstanding that they do not possess 
  the HOMO-LUMO gap. Technically we applied an artificial broadening 
  of molecular orbitals' eigenvalues by 400 K. Some of the even FSM numbers 
  allow an easily converging calculation, with the
  total energy and local moments which follow the general trend. Such results
  are added in the left and right panels of Fig.~\ref{fig:Fe3_FSM}.
\end{thebibliography}

\end{document}